\documentclass[a4paper,UKenglish,cleveref, autoref, thm-restate]{style}

\pdfoutput=1 
\hideLIPIcs  
\nolinenumbers

\renewcommand*{\orcidsymbol}  {}

\DeclareMathOperator{\argmax}{argmax}

\usepackage{amsmath,amsfonts, mathdots, amssymb, latexsym, amsbsy,bm,amsthm,mathtools}
\usepackage[vlined,ruled]{algorithm2e}
\usepackage[table]{xcolor}
\title{The Solidarity Cover Problem}

\author{Eran Rosenbluth}{RWTH-Aachen, Germany}{rosenbluth@informatik.rwth-aachen.de}{[]}{}
\funding{German Research Council (DFG), within Research Training Group 2236 (UnRAVeL)}

\authorrunning{E. Rosenbluth}

\Copyright{Eran Rosenbluth}

\ccsdesc[500]{Theory of computation~Problems, reductions and completeness}
\ccsdesc[500]{Theory of computation~Packing and covering problems}
\ccsdesc[500]{Theory of computation~Facility location and clustering}

\keywords{Domatic number, Domatic partition, Covering problems, Approximation algorithms, Hardness proofs}

\begin{document}
\maketitle

\begin{abstract}
Various real-world problems consist of partitioning a set of locations into disjoint subsets, each subset spread in a way that it covers the whole set with a certain radius. Given a finite set $S$, a metric $d$, and a radius $r$, define a subset $S'\subseteq S$ to be an \emph{$r$-cover} if and only if $\forall s\in S \;\exists s'\in S': d(s,s')\leq r$. We examine the problem of determining whether there exist $m$ disjoint $r$-covers, naming it the \emph{Solidarity Cover Problem} (SCP). We consider as well the related optimization problems of maximizing the number of $r$-covers, referred to as the \emph{partition size}, and minimizing the radius. We analyze the relation between the SCP and a graph problem known as the Domatic Number Problem (DNP), both hard problems in the general case. We show that the SCP is hard already in the Euclidean 2D setting, implying hardness of the DNP already in the \emph{unit disc graph} setting. As far as we know, the latter is a result yet to be shown. We use the tight approximation bound of $(1-o(1))/\ln(n)$
for the DNP's general case, shown by U.\,Feige, M.\,Halldórsson, G.\,Kortsarz, and A.\,Srinivasan (SIAM Journal on computing, 2002), to deduce the same bound for partition-size approximation of the SCP in the Euclidean space setting. We show an upper bound of $3$ and lower bounds of $2$ and $\sqrt{2}$ for approximating the minimal radius in different settings of the SCP. Lastly, in the Euclidean 2D setting we provide a general bicriteria-approximation scheme which allows a range of possibilities for trading the optimality of the radius in return for better approximation of the partition size and vice versa. We demonstrate a usage of the scheme which achieves an approximation of $(\frac{1}{16},2)$ for the partition size and radius respectively.
\end{abstract}
\section{Introduction}
Let $(S,d)$ be a metric space consisting of a finite set of points $S=\{s_1,\ldots,s_n\}$ and a metric ${d:S\times S\to \mathbb{R}_+}$, and let ${B_r(v)=\{u\in S\mid d(u,v)\le r\}}$ denote the set of elements in distance  $r\in\mathbb{R}$ from element $v\in S$. Given parameters $m\in[n],r\in\mathbb{R}$, the \emph{Solidarity Cover Problem} (SCP) is to decide whether there exist $m$ disjoint subsets of $S$ such that each subset covers $S$ with radius $r$, that is, whether there are ${S_1,\ldots,S_m,\ S_i\subseteq S,\ i\neq j \Rightarrow S_i\cap S_j=\emptyset}$\; such that ${\forall i\in [m] \bigcup_{v\in S_i} B_r(v)=S}$.
We will refer to such a partition as an \emph{$m$-solidarity-$r$-cover}, to a subset $S'\subseteq S$ that covers $S$ with radius $r$ as an \emph{$r$-cover}, and to a point that is covered in radius $r$ by $k$ disjoint subsets as being \emph{$k$-$r$-covered}. We refer to the number of subsets in the partition as the \emph{partition size}. Two optimization problems naturally follow the decision problem, maximizing the partition size given the radius and minimizing the radius given the partition size.

The problem resembles the already-studied \emph{Domatic Number Problem} (DNP). A \emph{dominating set} of an undirected graph $G=(V,E)$ is a subset $S'\subseteq V$ such that $\forall v\in V\; \exists w\in S' : \{v,w\}\in E$. Given an undirected graph $G=(V,E)$ and a parameter $m\in[|V|]$ the DNP is to decide whether there exist $m$ disjoint dominating sets of $V$, that is, whether there are  ${S_1,\ldots,S_m,S_i\subseteq V},\ i\neq j \Rightarrow S_i\cap S_j=\emptyset$\; such that $\forall i\in [m]\; S_i$ is a dominating set. The problem was first mentioned by this name in \cite{cockayne1975optimal}, was said to be NP-complete for $m\geq3$ in \cite[pg 190]{garey1979computers} without an explicit proof there, and one proof can be found in \cite{board2005texts}. The well-studied optimization version of the DNP is to maximize the partition size. For both, the SCP and the DNP, for $m=1$ the answer is trivially `yes'. For $m=2$ it is not difficult to verify that the answer is `yes' if and only if each element has at least one neighbor, that is,  another element within radius $r$ in the SCP and an adjacent vertex in the DNP.
The DNP is not difficult to reduce to the SCP's most general setting, and the SCP is straightforwardly reducible to the DNP. Yet, compared to the DNP the SCP allows the metric space to be something else than a graph with shortest edge-path distance function, and adds a radius parameter. These extensions give rise to questions about complexity in Euclidean space, most specifically in Euclidean 2D, and about approximability with regards to the radius. They match real-world motivations for the DNP e.g. network resources allocation \cite{pemmaraju2006energy} and facilities allocation \cite{feige2002approximating}. Another relevant scenario is sensing: Assume a set of locations for which a certain data needs to be repeatedly measured by sensors in those locations. Assume each sensor's reading is a good estimation in a certain radius, $r$, and there is a limitation on the frequency each sensor can be queried, $f_s$. Finally assume it is required to estimate the data in each location in frequency $f_r>f_s$. Let $m=\frac{f_r}{f_s}$, then an $m$-solidarity-$r$-cover of the set of sensors will allow having an estimation of the data in all locations in the required frequency by alternating between querying each of the $m$ covering subsets.

For the optimization version of the DNP, maximizing the partition size, a tight $(1-o(1))/\ln(n)$ approximation bound is known for general graphs \cite{feige2002approximating}. 
For interval graphs DNP, which coincide with Euclidean $1$-dimensional SCP, linear-time algorithms exist \cite{lu1990domatic}.
In the specific and interesting case of unit disc graphs a probabilistic constant factor approximation algorithm was presented in \cite{pandit2009approximation}, with a factor of over 500. 
However, no lower bound has been proven for that scenario. To the best of our knowledge, the DNP for unit disc graphs has not been even shown to be NP-hard.

Several well-studied covering problems are similar in some ways to the SCP but we found too different for their results to be of direct use in reasoning about the SCP. Such is the minimum set cover problem and such is the k-center problem. Some work on the latter though  \cite{gonzalez1985clustering} inspired the greedy technique used in \Cref{sec:approx_r}.

In \Cref{sec:decision_hardness} we prove NP-completeness of the SCP in Euclidean 2D space. This problem is equivalent to the DNP for unit disc graphs and so for the first time, as far as we know, the DNP for unit disc graphs is proven to be NP-hard.
In \Cref{sec:approx} we focus on approximability results. We examine the approximability of the SCP with regards to the partition size parameter. We show a relation between the SCP in Euclidean space and the DNP which implies the same tight approximation bound for the SCP as the one known for the DNP, $(1-o(1))/\ln(n)$ \cite{feige2002approximating}. In \Cref{sec:approx_r} we examine the approximability of the SCP with regards to the radius parameter. We introduce a 3-approximation algorithm 
generalizing the ideas in \cite{gonzalez1985clustering},  
and prove a 2-approximability lower bound
for the general metric space setting and a $\sqrt{2}$-approximability lower bound already for the Euclidean 2D setting.
In \Cref{sec:bicriteria} we introduce a bicriteria approximation scheme for the Euclidean 2D setting which allows trading the optimality of the radius in return for better approximation of the partition size and vice versa. We apply this scheme obtaining an exemplary $(\frac{1}{16},2)$-approximation algorithm for the partition size and radius respectively.

The following table summarizes our results for the SCP as well as results that are straightforwardly implied by previous results for the DNP. The latter are marked in grey. `L' and  `U' stand for lower and upper bound.

\begin{center} 
  \begin{tabular}{ c | c | c | c | c}	
	& Decision & Part' Size Approx' & Rad' Approx' & Bicriteria\\
	\hline
	General Metric Space & \cellcolor{gray!25}NP-Complete\cite{garey1979computers} & \cellcolor{gray!25}$(1-o(1))/\ln(n)$\cite{feige2002approximating} & L:$2$,\; U:$3$ & -\\
	\hline
	Euclidean Space & NP-Complete & $(1-o(1))/\ln(n)$ & L:$\sqrt{2}$,\; U:$3$ & -\\
	\hline
	Euclidean 2D Space & NP-Complete & \cellcolor{gray!25}O(1) \cite{pandit2009approximation} & L:$\sqrt{2}$,\; U:$3$ & 
	$\frac{1}{16}\ ,\ 2$

  \end{tabular}
\end{center}

\section{Decision Hardness}\label{sec:decision_hardness}
We start with showing that the SCP is NP-hard already in the Euclidean 2D setting, and as containment in NP is relatively straightforward completeness is proven. The hardness proof describes a reduction from the 3-coloring problem for a planar orthogonal graph $G=(V,E)$ drawn in an area of size $O(|V|^2)$, to the SCP. A graph is \emph{planar orthogonal} if and only if it is a planar graph whose edges are a combination of horizontal and vertical lines connecting integer-grid points. Any planar graph $G=(V,E)$ with degree at most $4$ admits a planar orthogonal embedding in an area of size $O(|V|^2)$ and that embedding can be computed in polynomial time \cite[Theorem 2]{valiant1981universality}. Hence, as the 3-coloring problem for planar graphs with degree at most 4 is known to be NP-hard \cite[Theorem 2.1]{garey1974some} so is the 3-coloring problem for planar orthogonal graphs drawn in an area of size $O(|V|^2)$.

\begin{lemma}\label{prop:3coloring2scp}
Let $G=(V,E)$\; be a graph such that:
\begin{itemize}
    \item[i.] $G$ is planar orthogonal.  
    \item[ii.] The area of $G$ in the plane is $O(|V|^2)$.
\end{itemize}
Then, there is a set of points $S\subseteq\mathbb{R}^2$ computable in polynomial time, such that: for every $1\leq r<\sqrt{2}$ \  there is a 3-solidarity-r-cover of $S$, if and only if $G$ is 3-colorable.
\end{lemma}

\begin{proof}
Please see Figure \ref{fig:graph_to_points_set} for an example of the following formal description. Let $G'=(V',E'), V'\subseteq \mathbb{Z}^2$\; be a scaling of $G$ such that each vertex in coordinates $(i,j)\in\mathbb{Z}^2$ is shifted to $(6i,6j)$ and the edges segments are lengthened accordingly. Also, assume ` $<$ '  is an arbitrary total ordering on $V'$ and, in accordance with (i), each edge $e_{uv}\in E'$ connecting  $u<v\in V'$ is defined as a sequence $e_{uv}=(e^{uv}_1=u,\ldots,e^{uv}_{k_{uv}}=v)\in(\mathbb{Z}^2)^{k_{uv}}$ of $k_{uv}\in \mathbb{N}$\; 1-step-afar grid points. Define the following subsets:
\begin{itemize}
\item For each $v=(v_x,v_y)\in V'$, if $deg(v)\geq 2$ then define $S^*_v=\{v\}$. Otherwise, if $v$ is isolated or has a single edge  connected to it from the right side then define $S^*_v= \{v,\ p^v_1=(v_x-1,v_y),\ p^v_2=(v_x-0.5,v_y+0.5)\}$, and otherwise define $S^*_v$ to contain $v$ and the equivalent two points considering the single edge direction.
\item Due to the scaling, for each edge $e_{uv}=(e^{uv}_1=u,\ldots,e^{uv}_{k_{uv}}=v)$, $e^{uv}_4$ must be either a part of a vertical segment or a horizontal segment but not both, that is, not a corner point. Assume w.l.o.g it is a part of a horizontal segment, then define 
\\$S^*_{uv}=\{{p^{uv}_{u}=(e^{uv}_{4_x}-0.5,e^{uv}_{4_y}+0.5),p^{uv}_{v}=(e^{uv}_{4_x}+0.5,e^{uv}_{4_y}+0.5),p^{uv}_{uv}=
(e^{uv}_{4_x},e^{uv}_{4_y}+1)\}}$.
\\The idea is that the three points in $S^*_{uv}$ are within radius 1 of each other, yet they are in distance $\geq\sqrt{2}$ from points that we do not want them to be in radius 1 of, as described later in the proof.
\item For each $e_{uv}\in E', e_{uv}=(e^{uv}_1=u,\ldots,e^{uv}_{k_{uv}}=v)$ define $S'_{uv}=\{e^{uv}_2,e^{uv}_3,e^{uv}_5,\ldots,e^{uv}_{k_{uv}-1}\}$, and $S_{uv}=S'_{uv}\cup S^*_{uv}$.
\end{itemize}

\begin{figure}[ht]
    \centering
    \makebox[\textwidth][c]{
    \includegraphics[width=0.95\textwidth]{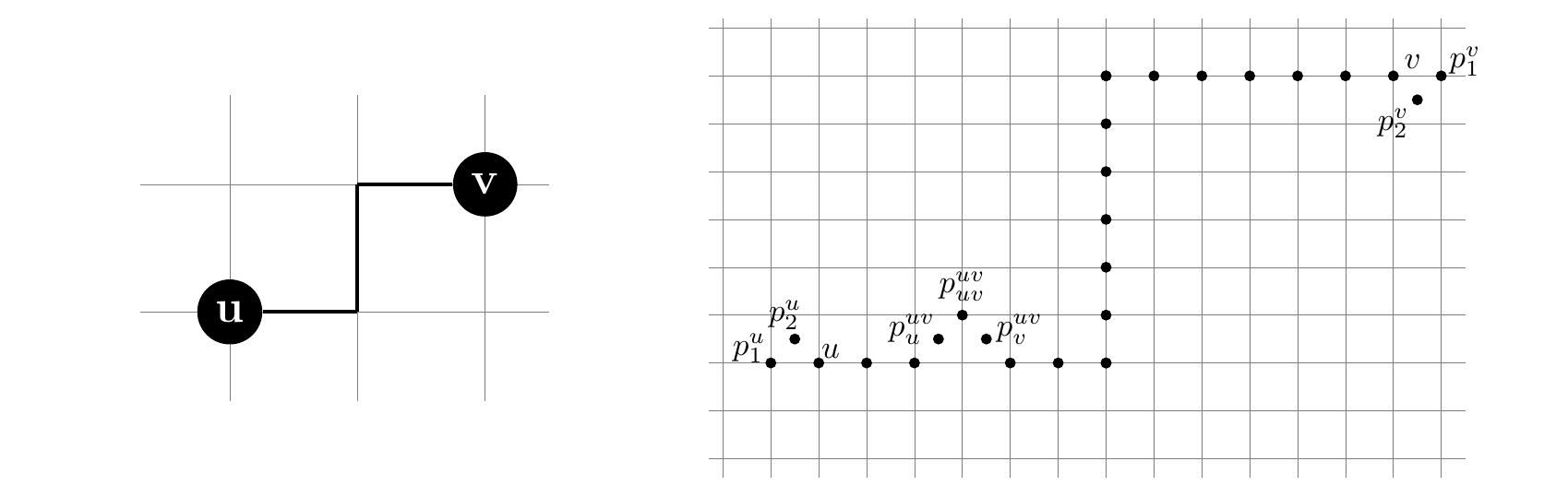}}
    \caption{An example of constructing $S$ given $G$. On the left is the  depiction of a planar-orthogonal graph $G$ consisting of two vertices $u,v$ and an edge between them. On the right is the depiction of the set of points $S$ constructed from $G$ assuming the vertices ordering $u<v$. In both diagrams the distance between subsequent grid-lines is 1.}
    \label{fig:graph_to_points_set}
\end{figure}

Finally, define $S=\{\bigcup_{v\in V'}S^*_v\} \cup \{\bigcup_{e_{uv}\in E'}S_{uv}\}$. Due to (ii) the described construction of $S$ can be executed in time polynomial in $\lvert{G}\rvert$. Due to (i), and the mentioned scaling, no points associated with an edge are in radius $< \sqrt{2}$ of points associated with another edge. The second important property of the construction is that for each edge $e_{uv}\in E'$, for each point $p\in S'_{uv}$\ ,\ $p$ has exactly two neighboring points within radius 1 in $S$. Moreover, for each two points in $S$ either they are within distance 1 of each other or they are at least $\sqrt{2}$ distance apart. We proceed to show the required relation between a $3$-solidarity-$1$-cover of $S$ and a 3-coloring of $G$, please see Figure \ref{fig:coloring_to_scf_example} for a demonstration of that relation.

For the first direction, assume $S_1,S_2,S_3\subseteq S$ form a 3-solidarity-1-cover of $S$. Let $e_{uv}\in E'$ and assume $u\in S_i$ and $v\in S_j$, we want to show that $i\neq j$. Note that necessarily $p^{uv}_{u}\in S_i$, $p^{uv}_{v}\in S_j$, and for each $4< k < k_{uv},\ (k-4)\mod 3=0$  it holds that $e^{uv}_k\in S_j$. These are due to the  radius-1 neighboring properties, and number of points in $e_{uv}$, guaranteed by the scaling of $G$ and the definition of $S$.
Finally, the only way for $p^{uv}_{uv}$ to be 3-1-covered is if $p^{u}_{uv}$ and $p^{v}_{uv}$ are assigned to different subsets, that is,  if $S_i\neq S_j$. Hence, coloring the vertices of $G'$ according to the solidarity-cover assignment of the vertices-points results in a valid 3-coloring, and since $G$ is just a spatial down-scaling of $G'$ the coloring is valid for it as well.

\begin{figure}[ht]
    \centering
    \makebox[\textwidth][c]{
    \includegraphics{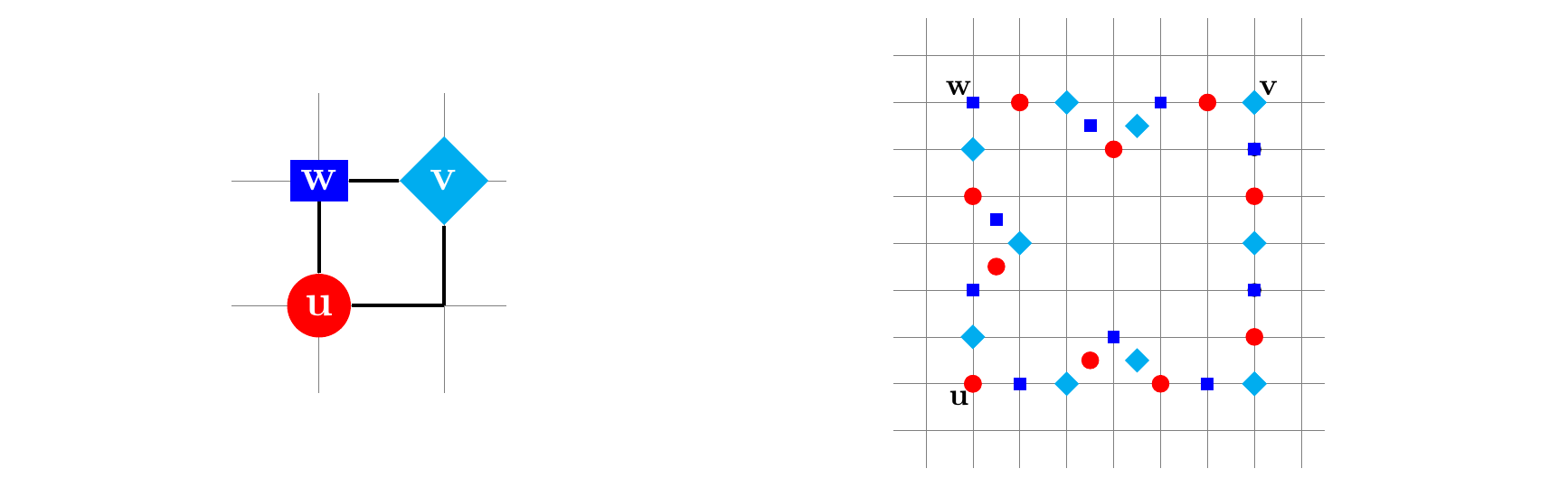}}
    \caption{A demonstration of the 3-coloring of a planar orthogonal graph $G$ and the corresponding 3-solidarity-1-cover of the constructed set of points $S$. On the left is a planar-orthogonal version of a 3-colored triangle graph $G$. The coloring is represented by 3 colors and 3 matching shapes. On the right is the set of points $S$ constructed from $G$ and partitioned into three disjoint subsets - represented by colors and shapes. The partition forms a 3-solidarity-1-cover. In both diagrams the distance between subsequent grid-lines is 1.}
    \label{fig:coloring_to_scf_example}
\end{figure}

For the second direction, assume $\tau:V\rightarrow \{1,2,3\}$ is a valid 3-coloring of $G$, then $\tau$ is a valid 3-coloring of $G'$. We define an assignment $\psi:S\rightarrow \{S_1,S_2,S_3\}$  of the points in $S$ to three subsets of $S$ as follows:
\begin{itemize}
\item  For each $v\in S$ such that $v\in V'$\; set  $\psi(v)=\tau(v)$. If $S^*_v$ contains two points in addition to $v$ then assign them to the remaining two subsets and get that each of the three points is 3-1-covered. Otherwise, there are at least two edges connected to $v$, assign the two first edge-points (which are not $v$) to the two remaining subsets and get that $v$ is 3-1-covered.
\item For each $e_{uv}\in E'$:
\begin{itemize}
    \item[a)] By assumption, and the definition of $\psi(w),w\in S$, we have $\psi(u)\neq \psi(v)$. Set $\psi(p^{uv}_{u})=\psi(u),\; \psi(p^{uv}_{v})=\psi(v)$, set $\psi(p^{uv}_{uv})$ to be the remaining subset, and get that each of the three points is 3-1-covered.
    \item[b)] 
    $\forall 4< k < (k_{uv}-3)
      \quad \psi(e^{uv}_k)= \begin{cases}
			\psi(v) & \text{(k-4)\ mod\ 3=0}
			\\S_q\neq \psi(v) & \text{(k-4)\ mod\ 3=1}
			\\S_r\neq \psi(v), \ r\neq q & \text{(k-4)\ mod\ 3=2}
		 \end{cases}
    $
    \\ Each of the above points has exactly two neighbors in radius 1, and the three have different assignments, hence each of them is 3-1-covered.
    \item[c)] Foreach $1<k<4$\; and \;$(k_{uv}-3) < k < k_{uv}$\; set $\psi(e^{uv}_k)$ to be a subset such that $e^{uv}_k,u,v$ are each 3-1-covered. It is quick to verify that such an assignment always exists given the definition of $S$ and of $\psi$ so far.
\end{itemize}
\end{itemize}
Overall, $\psi$ induces a 3-solidarity-1-cover of $S$.
\end{proof}
\begin{theorem}
The SCP is NP-Complete already for 3-solidarity-1-cover in Euclidean 2D space.
\end{theorem}

\begin{proof}
Hardness follows from \cref{prop:3coloring2scp} together with hardness of the 3-coloring problem for planar orthogonal graphs drawn in an area of size $O(|V|^2)$. Containment in NP follows from being able to use a partition description as a witness, verifiable in polynomial time, for the existence of an $m$-solidarity-$r$-cover.
\end{proof}

\section{Approximation Bounds}
\label{sec:approx}
In this section we examine approximation variants of the SCP. We begin with the objective of maximizing the partition size, proceed with the objective of minimizing the radius, and end with considering the two objectives simultaneously in the 2D setting.

\subsection{Partition Size Approximation}\label{sec:approx_m}
A tight bound of $(1-o(1))/\ln(n)$ approximation, assuming NP $\not\subseteq$  DTIME$(n^{O(\log\log(n))})$, was proven for the DNP in \cite{feige2002approximating}. 
We show that the same bound applies to the SCP in the Euclidean space setting.
\begin{lemma}\label{basic_m_approx_lemma}
Let $G=(V,E), V=\{v_1,\ldots,v_n\}$\; be an undirected graph and let $(S,d), S=\{s_1,\ldots,s_n\}, r\in\mathbb{R}$\; be a metric space and a radius such that $\{v_i,v_j\}\in E\Leftrightarrow d(s_i,s_j)\leq r$, then $G$ admits a domatic partition of size $m$ if and only if $S$ admits an $m$-solidarity-$r$-cover.
\end{lemma}
\begin{proof}
The assumption implies that for each $k\in[n],\{i_j\}_{j\in[k]}$ it holds that $\{v_{i_1},\ldots,v_{i_k}\}$ is a dominating set of $G$ if and only if $\{s_{i_1},\ldots,s_{i_k}\}$ is an r-cover of $S$. Also, trivially every two subsets of $V$ are disjoint if and only if their index-wise corresponding subsets of $S$ are disjoint. Hence, every domatic partition of $G$ of size $m$ corresponds to an $m$-solidarity-$r$-cover of $S$ and vice versa.
\end{proof}

\begin{theorem}
The SCP in Euclidean space adheres a tight bound of $(1-o(1))/\ln(n)$ for the partition size approximation, assuming \emph{NP} $\not\subseteq$ \emph{DTIME}$(n^{O(\log\log(n))})$.
\end{theorem}

\begin{proof}
For the lower bound we show a reduction from the DNP to the SCP in Euclidean space. Let $\rho$ denote the Euclidean distance function. Given $G=(V,E), V=\{v_1,\ldots,v_n\}$, we can compute in polynomial time a set $S=\{s_1,\ldots,s_n\}\subseteq\mathbb{R}^n$ and a radius $r\in\mathbb{R}$ such that $\rho(s_i,s_j)\leq r \Leftrightarrow \{v_i,v_j\}\in E$ \cite[Theorem 1]{maehara1984space}. By \cref{basic_m_approx_lemma} the above is a valid reduction.

For the upper bound, we can straightforwardly reduce the SCP to the DNP. Given a metric space $(S,d), S=\{s_1,\ldots,s_n\}, r\in\mathbb{R}$, define  $V=\{v_1,\ldots,v_n\},\ E=\{\{v_i,v_j\} \mid d(s_i,s_j)\leq r\}$. By \cref{basic_m_approx_lemma} the above is a valid reduction.
\end{proof}

\subsection{Radius Approximation}
\label{sec:approx_r}
Let $(S,d),|S|=n$\; be a finite metric space. For any given $m\in[n]$ there exists a radius $r\in \mathbb{R}$ for which there is an $m$-solidarity-$r$-cover since it is always possible to set $r=\max(\{d(s_1,s_2) \mid s_1,s_2\in S\})$.
Let $m\in[n]$ and let $r^*=\min(\{r\mid$ there exist an $m$-solidarity-$r$-cover$\})$. We describe a polynomial time algorithm that finds an $m$-solidarity-$3r^*$-cover. The main part of the algorithm is the subroutine $\textsc{GreedySC}$ (see pseudocode below) which is inspired by ideas in \cite{gonzalez1985clustering} where they are used for $k$-center clustering approximation. 
It receives as input the metric space, the partition size, and a radius $r$. If $r$ is a feasible radius for an $m$-solidarity-$r$-cover then $\textsc{GreedySC}$ outputs a partition, and if $\textsc{GreedySC}$ outputs a partition then it is an $m$-solidarity-$3r$-cover. 
Hence, let $$\hat{r}=\min(\{d(s_i,s_j)\ \mid s_i,s_j\in S, \textsc{GreedySC}((S,d),m,d(s_i,s_j))\neq\ \text{`false'}\}$$
then $\hat{r}\leq3r^*$. 
As the number of candidates for $\hat{r}$ is $|\{d(s_i,s_j)\ | s_i,s_j\in S|\leq\frac{1}{2}(n^2-n)$, and $\textsc{GreedySC}$ runs in polynomial time, we can find $\hat{r}$ and a corresponding partition in polynomial time.

\begin{algorithm}[htb]
\SetAlgorithmName{Algorithm}{alg:GreedySC}{Algorithm \textsc{GreedySC}}
	\caption{\textsc{GreedySC}}
	\KwIn{Finite metric space $(S,d)$, partition size $m\in[|S|]$, radius $r$}
	\KwOut{An $m$-solidarity-$3r$-cover or `false'}
	Initialize: Set $S_1,\ldots,S_m,P:=\emptyset$,  select  $p_1\in S$ arbitrarily, set $i:=1$, and $r_1:=\infty$\\
	\While{$r_i>2r$}{
    	Update $P:=P\cup\{p_{i}\}$\\
    	\If{P=S}{break;}
    	Select $p_{i+1}$ of maximal distance to all points in $P$ i.e.
    	$$p_{i+1}\in \argmax_{p\in (S\setminus{P})}\left \{\min_{v\in P} d(v,p)\right\}$$\\
    	Set $r_{i+1}:=\min_{v\in P} d(v,p_{i+1})$;
    	\\Set $i:=i+1$;
	}
	Set $i:=i-1$\\
	\For{$k=1$ to $i$}{
	Let $B_{p_k}(r)$ denote the points in the radius $r$ ball  centered around $p_k$\\
	\If{$|B_{p_k}(r)|<m$} {
	\Return{false}
	}
	Assign the points in $B_{p_k}(r)$ to all $m$ different subsets $S_1,\ldots,S_m$
	}
	Assign the points in ($S\setminus{\bigcup_{k=1}^i B_{p_k}(r)}$) arbitrarily to $S_1,\ldots,S_m$\\
 \Return{$\{S_1, \ldots, S_m\}$}
\end{algorithm}
 
\begin{lemma}\label{lemma_3_approx}
Let $(S,d), |S|=n$ be a finite metric space, $m\in[n], r\in\mathbb{R}$, then:
\begin{itemize}
    \item[1)] If $S$ admits an $m$-solidarity-$r$-cover, then \textsc{GreedySC}$((S,d),m,r)$ will return a partition.
    \item[2)] If \textsc{GreedySC}$((S,d),m,r)$ returns a partition, then that partition is an $m$-solidarity-$3r$-cover.
\end{itemize}
\end{lemma}

\begin{proof}
Note that when referring to variables in the subroutine i.e. the set variables, they are considered in their final state - before the subroutine terminates.

First, we show that if there is an $m$-solidarity-$r$-cover then the subroutine will not terminate with `false'. Assume otherwise and let w.l.o.g. $|B_{p_1}|<m$, implying that there are no $m$ points in $S$ which are within radius $r$ of $p_1$. Hence, there cannot be $m$ disjoint subsets that cover $p_1$ with radius $r$, in contradiction to the existence of an $m$-solidarity-$r$-cover.

Next, we show that $(S_1, \ldots, S_m)$ are disjoint. Assume otherwise and let $p\in S$ such that $p\in S_j,p\in S_k,j\not = k$. By the second part of the subroutine, which assigns points to subsets, necessarily $p\in B_{p_j}(r), p\in B_{p_k}(r)$, implying $d(p_j,p_k)\le 2r$, in contradiction to the while-condition in the subroutine's first part.

Finally, we show that each $S_i$ is a $3r$-cover. Assume otherwise, w.l.o.g. assume $S_1$ is not a $3r$-cover and let $w\in S$ such that $\forall q\in S_1\; d(w,q)>3r$. 
By the second part of the subroutine, which assigns points to subsets, $w\not\in P$ and also $\forall u\in P\; \exists v\in S_1 : d(u,v)\leq r$. Hence, necessarily $\forall u\in P \;\; d(w,u)>2r$ in contradiction to the termination of the subroutine without adding $w$ to $P$.
\end{proof}

\begin{theorem}
Let $(S,d),|S|=n$\; be a finite metric space, let $m\in[n]$, and let $r^*=\min(\{r\mid$ there exist an $m$-solidarity-$r$-cover$\})$. Then, an $m$-solidarity-$3r^*$-cover can be found in polynomial time.
\end{theorem}
\begin{proof}
There are $O(|S|^2)$ possible radii thus we can search in polynomial time for the minimal radius for which \emph{GreedySC} returns a partition. By \Cref{lemma_3_approx} necessarily that partition is an $m$-solidarity-$3r^*$-cover.
\end{proof}

\begin{theorem}\label{thm:2 approx lower bound}
There does not exist a polynomial-time c-approximation of the radius, with $c<2$, for the SCP in general metric space setting, unless \emph{P}$=$\emph{NP}.
\end{theorem}

\begin{proof}
We show that otherwise the DNP can be solved in polynomial time, in contradiction to it being NP-hard. Assume by contradiction that there exists a c-approximation algorithm $A$ for some $c<2$.
Given a DNP instance $G=(V,E),\ V=\{v_1,\ldots,v_n\}, m\in[n]$, we can construct in polynomial time a metric space $(S,d), S=\{p_1,\ldots,p_n\}$, $d(p_i,p_j):=\min(\{|P| \mid P\subseteq E \mbox{ is a } v_i-v_j \mbox{ path in }G\})$, that is, the distance between two points is defined to be the length (in edge-count) of the shortest-path between their corresponding vertices in the graph. It is clear that there is a domatic partition of size $m$ for $G$ if and only if there is an $m$-solidarity-$1$-cover for $S$ or, in other words, the minimal feasible radius for a solidarity cover with parameter $m$ is 1. Hence, if there is a domatic partition then $A$ must return "$r<2$" and the result will indicate that actually the minimal feasible radius is $1$ as due to our definition of $d$ the potential minimal feasible values are discrete. In the other direction, if the approximation algorithm returns "$r<2$" then by the same reasoning it indicates the existence of a domatic partition of size $m$ for $G$.
\end{proof}

\begin{theorem}\label{thm:sqrt2 approx lower bound}
There does not exist a polynomial-time c-approximation of the radius, $c<\sqrt{2}$, for the SCP in Euclidean space setting of dimension $n\geq2$, unless \emph{P}$=$\emph{NP}.
\end{theorem}

\begin{proof}
We show that otherwise the 3-coloring problem for planar orthogonal graph $G=(V,E)$ drawn in an $O(|V|^2)$ area can be solved in polynomial time, in contradiction to it being NP-hard.
According to \cref{prop:3coloring2scp}, given an instance of the mentioned coloring problem we can construct a set $S\subseteq \mathbb{R}^2$ that admits a 3-solidarity-r-cover  for every $1\leq r<\sqrt{2}$, if and only if a 3-coloring exists. Assume there is a c-approximation algorithm, $c<\sqrt{2}$, then given the set of points and partition size $3$ it will return a 3-solidarity-r-cover with $r<\sqrt{2}$ if and only if a 3-coloring exists for the graph.
\end{proof}

\subsection{Bicriterial Approximation in the Euclidean 2D Setting}
\label{sec:bicriteria}
In the spirit of \cite[Section 3.1]{pandit2009approximation} we show a deterministic bicriteria approximation scheme for the Euclidean 2D setting, allowing to improve the approximability of the radius on account of the optimality of the partition size, and vice versa. The scheme enables a range of trade-offs - rather than only one specific -  between the two approximation factors. Throughout this section, we assume the distance function, $d$, of our metric space to be the Euclidean distance.
Similarly to \Cref{sec:approx_r} we rely on the polynomial size of the set of radii to be considered.
Let $S\subset\mathbb{R}^2,|S|=n$\; be a finite set of points in the plane. Let $m\in[n]$ and let $r^*=\min(\{r\mid$ there exist an $m$-solidarity-$r$-cover$\})$. Assume a plane that is divided into squares of diameter $r'$ and let $f(r,r')$ be the maximal number of squares intersecting a circle of radius $r$ placed anywhere in that plane. We describe a polynomial time algorithm which, given a desired radius approximation factor $\beta$, finds an $\frac{1}{f(r,(\beta -1)r)}m$-solidarity-$\beta r^*$-cover.
The main part of the algorithm is the subroutine $\textsc{SquaresSC}$ (see pseudocode below). It receives as input the set of points, the partition size, the desired radius approximation factor $\beta$, and a radius $r$. If an $m$-solidarity-$r$-cover exists then $\textsc{SquaresSC}$ outputs a partition, and if $\textsc{SquaresSC}$ outputs a partition then that partition is $\frac{1}{f(r,(\beta -1)r)}m$-solidarity-$\beta r$-cover. Hence, let 
$${\hat{r}=\min(\{d(s_i,s_j) \mid  s_i,s_j\in S,\ \textsc{SquaresSC}((S,d),m,\beta,d(s_i,s_j))\neq\ \text{`false'}\}})$$
then $\hat{r}\leq\beta r^*$. 
As the number of candidates for $\hat{r}$ is $|\{d(s_i,s_j)\ | s_i,s_j\in S|\leq\frac{1}{2}(n^2-n)$, and $\textsc{SquaresSC}$ runs in polynomial time, we can find $\hat{r}$ and a corresponding partition in polynomial time.
\begin{algorithm}[htb]
\SetAlgorithmName{Algorithm}{alg:SquaresSC}{Algorithm \textsc{SquaresSC}}
	\caption{\textsc{SquaresSC}}
	\KwIn{Finite set of points in the plane $S\subset\mathbb{R}^2$, partition size $m\in[|S|]$, radius approximation factor $1<\beta\in\mathbb{R}$, radius $r\in\mathbb{R}$}
	\KwOut{A $\frac{1}{f(r,(\beta -1)r)} m$-solidarity-$\beta r$-cover or `false'}
	Set $m'=\frac{1}{f(r,(\beta -1)r)} m$, $S_1,\ldots,S_{m'}:=\emptyset$;\\
	Let $(t,l,b,\rho)$ be the top, left, bottom, and right extreme coordinates of points in $S$;\\ 
	Split the rectangle with corners
	$(t,l),(t,\rho),(b,\rho),(b,l)$
	to squares of diameter $(\beta-1)r$;\\
	\ForEach{square $s$}{
	\If{$s$ has at least $m'$ points}{assign the points in $s$ such that each of the $m'$ subsets is assigned at least one point;}
	\Else{
	assign the points in $s$ to the $m'$ subsets arbitrarily;}
	}
	\If{$\{S_1,\ldots,S_{m'}\}$ 
	is an $m'$-solidarity-$\beta r$-cover} 
	{\Return{ $S_1,\ldots,S_{m'}$}}
	\Else{\Return{`false'}}
\end{algorithm}

\begin{lemma}
\label{thm:bicreteria_approx}
Let $S\subseteq\mathbb{R}^2,|S|=n$\; be a finite set of points in the plane, $m\in[n], r\in\mathbb{R}, 1<\beta\in\mathbb{R}$, then:
\begin{itemize}
    \item[1)] If $S$ admits an $m$-solidarity-$r$-cover, then \textsc{SquaresSC}$((S,d),m,\beta,r)$ will return a partition.
    \item[2)] If \textsc{SquaresSC}$((S,d),m,\beta,r)$ returns a partition, then that partition is a $\frac{1}{f(r,(\beta -1)r)}m$-solidarity-$\beta r$-cover.
\end{itemize}
\end{lemma}
\begin{proof}
Assume $S$ admits an $m$-solidarity-$r$-cover. Let $S_1,\ldots,S_{m'}$ be the partition constructed before the last line of \textsc{SquaresSC}, and let $p\in S$. Then, assuming any division of the plane into squares of diameter $(\beta -1)r$, the circle of radius $r$ around $p$ necessarily intersects with a square that contains at least $\frac{1}{f(r,(\beta -1)r)} m$ points. Hence, by the first "For each" statement of the subroutine, each subset $S_i, i\in [m']$ has a point that belongs to that square and so it is at most $r+(\beta -1)r=\beta r$\; distant from $p$.

By the last line of the subroutine, if the constructed partition is not an $\frac{1}{f(r,(\beta -1)r)}m$-solidarity-$\beta r$-cover then `false' will be returned.
\end{proof}

\begin{lemma}
\label{lem:sixteen_two_approx}
Assume a plane that is divided to squares of diameter $r$,  then the maximal number of squares intersecting a circle of radius $r$ placed on that plane is exactly 16.
\end{lemma}
\begin{proof}
To show that 16 is an upper bound, instead of looking on a circle of radius $r$ we look on a square (parallel to the axis) with side-length $2r$, that is, big enough to contain the circle. The side length of the grid-square is $\frac{r}{\sqrt{2}}$, hence, the maximum number of grid lines it can intersect in each dimension is 4, hence, the maximum number of grid-squares it can intersect is 16.
To show that the maximal number of intersecting squares is at least 16 we look for example on the circle of radius $r$ around the origin. Assume the grid is aligned with the origin and the top left coordinates of square $Q_{ij}$ are 
 $(i\frac{r}{\sqrt{2}},(j+1)\frac{r}{\sqrt{2}})$, it is easy to verify that the circle intersects squares $\{Q_{i,j}\}_{i,j\in\{ -2, -1 , 0, 1 \}}$.
\end{proof}
\begin{theorem}
Given a partition size $m$, let $r^*=\min(\{r\mid$ there exist an $m$-solidarity-$r$-cover$\})$. Then, a $\frac{1}{16}$m-solidarity-$2r^*$-cover can be found in polynomial time.
\end{theorem}
\begin{proof}
There are $O(|S|^2)$ possible radii thus we can search in polynomial time for the minimal radius for which \emph{SquaresSC} with $\beta=2$ returns a partition. By \Cref{thm:bicreteria_approx} and \Cref{lem:sixteen_two_approx} necessarily that partition is a $\frac{1}{16}$m-solidarity-$2r^*$.
\end{proof}

\section{Conclusion and Outlook}
We have seen that the SCP is a generalization of the DNP also in the Euclidean space setting, and bears the same decision and partition size approximability hardness. Already for the very specific but important Euclidean 2D setting, in which it coincides with the DNP for unit disc graphs, the SCP is hard to decide. Still, the SCP brings news with it. Having the radius parameter opens the possibility for an optimal partition size in return for a 3-approximation of the radius. It is possible that the approximability can be improved but only down to a factor of 2 in the general case and $\sqrt{2}$ in more specific settings. Alternatively, in the Euclidean 2D setting both optimalities can be simultaneously compromised, improving the approximation of the radius - potentially even beyond the single-criteria lower bound - while degrading the approximation of the partition size.

Several questions naturally follow our investigation of the SCP:
\\\textbf{Radius Approximation Bound.} The gap between the radius approximability bounds may be narrowed.
\\\textbf{Bicriteria Approximation.} The trade-off between radius-approximability and partition size approximability may be improved. Also, bicriteria hardness results may be proved.
\\\textbf{Redundancy Extension.} We have defined a point $p\in S$ to be covered by a subset $S_i\subseteq S$ if the subset contains at least one point close to it, $p'\in S_i,\ d(p,p')\leq r$. In various potential applications, e.g. sensing, it is beneficial to have a coverage redundancy i.e. to consider a point $p\in S$ as covered by a subset $S_i\subseteq S$ only if the subset contains $\ell>1$ points close to it, that is, if there exist $p_1,\ldots,p_{\ell}\in S_i,\ \forall j\in[\ell]\; d(p,p_j)\leq r$. Obviously the redundancy version is at least as hard but it raises the question whether it is also at most as hard i.e. whether it has the same approximation possibilities.
\\\textbf{Outliers.} We have considered a partition to be a cover only if every point is covered by every subset. In the somewhat resembling $k$-center problem it was observed that by allowing a small fraction of outliers the radius could be significantly reduced. One could think then that in the SCP requiring only a high fraction of the points - and not all of them - to be covered by every subset will allow finding a solidarity cover with smaller radius in polynomial time.
\bibliographystyle{plainurl}
\bibliography{literature}

\end{document}